\documentclass{aa}
\usepackage{graphics,epsfig}
\usepackage{natbib}
\bibpunct{(}{)}{;}{a}{}{;}
\def\Myr{$M_\odot$~yr$^{-1}$}
\def\etaR{\eta_{\rm R}}
\def\spose#1{\hbox to 0pt{#1\hss}}
\def\lta{\mathrel{\spose{\lower 3pt\hbox{$\mathchar"218$}}
     \raise 2.0pt\hbox{$\mathchar"13C$}}}
\def\gta{\mathrel{\spose{\lower 3pt\hbox{$\mathchar"218$}}
     \raise 2.0pt\hbox{$\mathchar"13E$}}}

\begin{document}

\titlerunning{Test of empirical mass loss formulas}

\title{A critical test of empirical mass loss formulas  \\
       applied to individual giants and supergiants}

\author{K.-P.~Schr\"oder\inst{1,2} \and M.~Cuntz\inst{3,4} }


\offprints{K.-P.~Schr\"oder}

\institute{
       Departamento de Astronom{\'{\i}}a, A. P. 144, Universidad de Guanajuato,
       36000 Guanajuato, GTO, Mexico; \\
       \email{kps@astro.ugto.mx}
\and
       Zentrum f\"ur Astronomie und Astrophysik, Technische Universit\"at 
       Berlin, Hardenbergstr. 36, 10623 Berlin, Germany  
\and
       Department of Physics, Science Hall, University of Texas at Arlington,
       Arlington, TX 76019-0059, USA; \\
       \email{cuntz@uta.edu}
\and
       Institut f\"ur Theoretische Astrophysik, Universit\"at Heidelberg,
       Albert \"Uberle Str. 2, 69120 Heidelberg, Germany
} 

   \date{Received October 27, 2006; accepted January 10, 2007}

   \abstract{
To test our new, improved Reimers-type mass-loss relation, given by 
Schr\"oder \& Cuntz in 2005, we take a look at the best studied 
galactic giants and supergiants
--- particularly those with spatially resolved circumstellar shells and winds,
obtained directly or by means of a companion acting as a probing light source.
Together with well-known physical parameters, the selected stars provide the 
most powerful and critical observational venues for assessing the validity of
parameterized mass-loss relations for cool winds not driven by molecules or 
dust. In this study, star by star, we compare our previously published 
relation with the original Reimers relation (1975), the Lamers relation 
(1981), and the two relations by de Jager and his group (1988, 1990).  
The input data,
especially the stellar masses, have been constrained using detailed stellar
evolution models.  We find that only the relationship by Schr\"oder \& Cuntz
agrees, within the error bars, with the observed mass-loss rates for all giants
and supergiants.
}  \maketitle

\section{Introduction}

The most common type of mass loss of stars evolved away from the main-sequence
is a ``cool wind'' as found for most red giants and supergiants of spectral 
type K and early M.  However, despite several decades of research, the
processes driving these winds are still poorly understood.  These stars do
not produce dust in sufficient quantities and have insufficient CS opacity
to facilitate significant radiation pressure for mass loss --- see, e.g.,
Schr\"oder et al.~(2003) for a model-based description of this kind of
mass loss.  Hence, the energy reservoir for driving these cool winds is
most likely related to some type of turbulent energy density, within the
chromosphere or underneath, possibly associated with the manifestation of
(magneto-)acoustic waves (e.g., Musielak 2004, and references therein).
While just a small fraction of this energy is required for initiating
cool star winds, many details on the generation of non-radiative energy
as well as the acceleration of winds in the upper chromospheric regions
are still unsolved.

In the absence of any physical model of cool wind driving processes, 
the (semi-)empirical, so-called ``Reimers law" or ``Reimers formula"
(Reimers 1975) has been the popular choice of a parameterized mass-loss 
description for cool winds not driven by dust since over 30 years.
Albeit making a rather adequate representation of observed mass-loss
rates over a broad range, i.e., between on the order of $10^{-9}$
and $10^{-6} M_{\odot}$~yr$^{-1}$, the original Reimers relation does 
not provide any realistic physical reasoning on how the mass loss is
generated.  This has fueled our motivation for an attempt to derive the
Reimers relation based on a physical approach: it considers a detailed
assessment of the energy reservoir of the turbulent chromosphere with
particular emphasis on the action of Alfv\'en waves (Schr\"oder \& Cuntz
2005; Paper~I).  By doing so, we arrived at an improved relationship
now of physical reasoning, which also overcame the most significant
shortcoming of the original Reimers law, i.e., the variable fitting
factor $\etaR$.

A further motivation for an improved mass-loss description arises from
the need for an accurate theoretical initial-mass / final-mass relation
for stellar evolution models.  This is, e.g., a pivotal requirement for
a quantitative model of the galactic white dwarf population
(e.g., Schr\"oder et al. 2004), including their cooling times, an
excellent galactic disk chronometer.  Other applications include 
evolved stars in the stellar population, particularly their mass-loss
histories (e.g., Schr\"oder et al. 1999; Schr\"oder \& Sedlmayr 2001;
Schr\"oder \& Pagel 2003).

In Schr\"oder \& Cuntz (2005), we re-derive the classic Reimers relation
(see Sect. 2.1).  However, within this derivation, two extra terms of
moderate impact on the empirical mass-loss prediction arise:
one dependent on the stellar effective temperature $T_{\rm eff}$ and
the other on the stellar surface gravitational acceleration $g_*$,
with the final result given as
\begin{equation}
\dot{M} \ = \ \eta \cdot \frac{L_* R_*}{M_*} \cdot
{\Bigl(\frac{T_{\rm eff}}{4000~{\rm K}}\Bigr)}^{3.5} \cdot
\Bigl(1+\frac{g_{\odot}}{4300 \cdot g_*}\Bigr)
\end{equation}
with $\eta = 8 \cdot 10^{-14} M_\odot$ yr$^{-1}$, $g_\odot$
as solar surface gravitational acceleration, and
$L_*$, $R_*$, and $M_*$ in solar units.

Herein, the two main points of consideration are the following.
First, the surface-integrated mechanical energy flux at the bottom of the
chromosphere is found to be proportional to $T_{\rm eff}^{7.5}$.  After
a substitution of the stellar luminosity $L$, which is proportional to
$T_{\rm eff}^4$, a temperature-dependent term alike $T_{\rm eff}^{3.5}$
is obtained.  Though subtle, we found evidence for this extra term
from the masses lost on the RGB by globular cluster stars of
very different metallicity, as sensitively revealed by the colors of
their present-day horizontal branch stars (see Schr\"oder \& Cuntz 2005).

Unfortunately, even the best-studied individual giants discussed below 
still come with uncertainties of their physical parameters too
large, and a relative range of $T_{\rm eff}$ too small, to allow a direct 
assessment of the $T_{\rm eff}$-exponent. Furthermore, different trial 
versions would require recalibrated $\eta$-values, which reduce the
resulting difference in predicted mass loss.

The second extra term stems from the increase in characteristic
chromospheric extent as function of decreasing stellar surface gravity,
after substituting it by the stellar radius $R_*$,
and it does make a noticeable difference with most stars studied here.
Examples of large chromospheric extent include M-type supergiants
(Hartmann \& Avrett 1984; Airapetian et al. 2000).  Considering
the upper chromosphere as a starting point of the stellar wind, the
implication is that a relatively low amount of potential energy is
required to initiate the wind, if it starts further away from
the photosphere. This term can make up to a factor of 2 to 3
for supergiants on the verge of a dust-driven wind, albeit not fully 
developed, as found in $\alpha$~Ori (M2~Iab), see Sect. 3.6.

Therefore, in the following, we make use of the best-studied galactic
giants and supergiants, both in terms of stellar parameters and
observed mass loss, to critically test our improved Reimers-type
mass-loss relation.  Our aim is to compare its predictions with
the results of other mass-loss formulas given in the literature.
The by far most accurate, empirical mass-loss determinations arise
from a small number of giants and supergiants for which direct or
indirect, as by means of an orbiting companion acting as a light probe,
{\it spatial resolution of the circumstellar envelope} exists.
This criterion yields a the sample of case studies presented below,
with the exception of the old and very near RGB giant $\alpha$~Boo.

In Sect. 2, we present the other mass loss formulas from the
literature used for this study, as well as information of the
computation of the stellar evolution models and the statistical
method to establish the uncertainty bars of the predicted mass loss
rates.  Section 3 lists our case studies and Sect. 4 presents our
conclusions.


\section{Approach}

\subsection{Empirical mass-loss formulas}

For clarity, we concentrate on comparing our new relation only with the 
few, most commonly used mass-loss relations for cool winds. 
The popular, original ``Reimers law",
\begin{equation}
\dot{M} \ = \ \etaR \cdot \frac{L_* R_*}{M_*} 
\end{equation}
has been conceived already over 30 years ago (Reimers 1975) [R75], in a time
when the physical processes governing the initiation of mass loss in cool,
low-gravity stars were virtually unknown.  Consequently, Reimers never claimed
any physical reason for this relation, as he rather based it on pure
dimensional arguments.  Kudritzki \& Reimers (1978) later on calibrated
$\etaR$, the only free parameter, based on observed mass-loss rates of three
well-studied M-type supergiants, i.e., $\alpha$~Sco, $\alpha^1$~Her, and
$\delta^2$~Lyr, deducing a value of $\etaR = 5.5 \cdot 10^{-13}$.  However,
the wide-spread application of the ``Reimers law", over the years, 
has seen the usage of a range of different values for $\etaR$ to fit
mass-loss rates for a large variety of stars, which significantly
diminished the significance of the proposed relation.  In fact, the
most commonly used value has been $\etaR = 2 \cdot 10^{-13}$, which
will be considered in the forthcoming comparisons.

Another early, purely empirical relation for was suggested by Lamers 
(1981) [L81] for O and B giants, which in terms of $L_*$, $R_*$, and
$M_*$ (in solar units) reads
\begin{equation}
\dot{M} \ = \ 10^{-4.83} \cdot \left( {L_* \over 10^6} \right)^{1.42} \cdot
                               \left( {R_* \over 10^3} \right)^{0.61} \cdot
                               \left( {M_* \over 10^0} \right)^{-0.99}  \ .
\end{equation}
While clearly not intended by the author for describing the mass loss of
late-type giants and supergiants, it has nonetheless previously been used for
this purpose.  Therefore, we have included it for comparison.

De~Jager et al. (1988) [dJNH88] presented an empirical formula based on
Chebychev polynomials by which they obtained the best match of the mass-loss
rates for a large sample of stars.  In this case, the predicted mass-loss
rates were given as functions of $L_*$ and $T_{\rm eff}$.  In a subsequent
paper, Nieuwenhuijzen \& de~Jager (1990) [NdJ90] presented a simpler
parameterization to fit their updated sample, which again included stars of
high luminosity.  They found
\begin{equation}
\dot{M} \ = \ 9.63 \cdot 10^{-15} \cdot L_*^{1.42} \cdot R_*^{0.81}
                                                   \cdot M_*^{0.16}  \ .
\end{equation}

However, from todays perspective, it is noteworthy that this type of
approach does not distinguish between the two main groups of cool giants 
and supergiants with very different mass-loss mechanisms, i.e., highly
evolved cool supergiants with dust-driven winds, and less evolved giants
and supergiants with cool winds {\it not} driven by radiation pressure on 
dust.  This phenomenon results in an almost discontinuous change of the stellar
mass-loss behavior as function of stellar luminosity, as e.g. also found
in the empirical analysis by Judge \& Stencel (1991).  Furthermore,
as a trade-off for their very large sample size, the individual stellar
properties often lack precision.  Additionally, these relationships were
based on mass loss and luminosity determinations obtained prior to
new developments such a improved angular diameter measurements, new
stellar spectral analyses, and parallax determinations by Hipparcos.

\subsection{Stellar evolution models}

In order to complete the set of physical parameters, especially for
deriving the stellar parameters $L$ and $T_{\rm eff}$ of the giants and
supergiants studied here, we have computed evolution tracks for finding 
the best-matching masses.  For this purpose, we use a fast and well-tested 
evolution code (Eggleton 1971; Pols et al. 1997) in combination with 
a mass-loss description. 

The evolution code and its semi-empirical choice of convection parameters 
has been tested very sensitively by means of supergiants in eclipsing 
spectroscopic binaries with well-known physical parameters (Schr\"oder et al.
1997) allowing to deduce quantitatively reliable models for evolved stars.
Furthermore, the evolution code readily accepts any mass-loss prescription
as part of its boundary condition at each individual time-step (see Schr\"oder 
et al. 1999).  This is an ideal property for devising stellar evolutionary
computations including the associated total mass loss (e.g., Schr\"oder \&
Sedlmayr 2001; Schr\"oder \& Cuntz 2005).  At the same time, a connection is
obtained between the present-day mass and the initial mass for any highly
evolved star with a cool wind, as e.g. undertaken for $\alpha^1$~Her
(see Sect. 3.5).

Most important for this work is, however, that the use of this 
well-calibrated evolution code helps us to test and further reduce
uncertainties in the physical properties of the stars studied here.
This is pivotal for assessing the various empirical mass-loss relationships, 
because rarely are there direct empirical constraints on (super-)giant's 
masses. Also, when available, conflicting empirical constraints can leave 
us with alternative parameter sets.

\subsection{Uncertainty bar analysis}

An important aspect of any meaningful comparison between predicted mass-loss
rates and observations or between themselves is the determination of
uncertainty bars.  In case of predicted mass-loss rates, these uncertainties
evidently depend on the uncertainties of the input parameters.  The various
mass-loss relationships considered here constitute different function
with some differences concerning their input parameters.  Specifically, the
mass-loss relations by Reimers (1975), Lamers (1981), and Nieuwenhuijzen \&
de~Jager (1990) depend on $L_*$, $R_*$, and $M_*$ (see Eqs. 2, 3, and 4), the
relation by de~Jager et al. (1988) depends on $L_*$ and $T_{\rm eff}$, and
the new relationship by Schr\"oder \& Cuntz (2005) depends on $L_*$, $R_*$,
$T_{\rm eff}$, and $M_*$ (see Eq. 1).

The uncertainty bars of these input quantities for the different stars, as well
as the input quantities themselves, are carefully evaluated in Sect. 3, based 
on detailed observations and measurements.  In that respect, it should be noted
that out of $L_*$, $R_*$, and $T_{\rm eff}$, only two variables are physically
independent of each other.  Therefore, the remaining parameter, if required, 
as well as its uncertainty bar, needs to be calculated from the two other 
parameters available.  Furthermore, the surface gravity, as well as its 
uncertainty, used in the relation by Schr\"oder \& Cuntz (2005), 
is also not independent, but given by $R_*$ and $M_*$.

In order to obtain the uncertainty of each predicted mass-loss rate for each 
star, we take the following approach: by utilizing a random number generator, 
we produce a random series of mass-loss rates $X_{ij}$ for each case~$i$ (i.e.,
star or predicted mass-loss rate) by statistically varying the input data using
a Gaussian distribution.  The mean and the standard deviation for each input
quantity are given by observations (see Sect. 3).  The uncertainty bar for
each case ${\delta}X_{i}$ is then defined as
\begin{equation}
{\delta}X_i \ = \ \sqrt{ \sum^n_{j=1}{(X_{ij}-\bar{X_i})^2 } \over {n-1} }
\end{equation}
with $\bar{X_i}$ as predicted mass loss rate and $n$ as number of data
for each case $i$.
The random quantities $X_{ij}$ were calculated by means of the standard 
routines {\tt RAN1} and {\tt GASDEV} (Press et al. 1986):  {\tt RAN1}
delivers a uniform deviate of random numbers, which is used as input for
{\tt GASDEV}.  {\tt GASDEV}  provides a normal deviate for a specified
mean value and standard deviation.  For each case, we produced a series
of 50 runs.


  \begin{figure}
  \vspace{7cm}
  \includegraphics{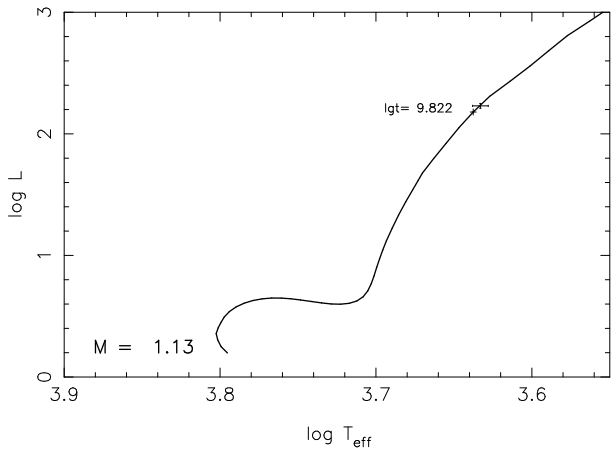}
  \caption{Luminosity versus effective temperature for stellar evolution
  computations of Arcturus.  Note that the observationally deduced value of
  $T_{\rm eff}$ is very well matched for an evolution track of $1.13 M_{\odot}$ with
  subsolar metallicity of $Z = 0.01$, implying an age of $6.6 \cdot 10^9$ yr.}
  \end{figure}

\begin{table*}[bht]
  \caption{Recently published physical parameters of $\alpha$~Boo.}

  \begin{center}
    \leavevmode
    \footnotesize
    \begin{tabular}{l c c c c c c c}
\hline \hline
\noalign{\vspace{0.03in}}
Source  & $d$ & $BC+A_v$ & $\log{L_*/L_{\odot}}$ & $T_{\rm eff}$ & $R_*$ & 
     $M_*$ & ${\dot{M}_{\rm obs}}$ \\
\omit  & [pc]& \omit & \omit & [K] & [$R_{\odot}$] & 
         [$M_{\odot}$]  & [{\Myr}] \\
\noalign{\vspace{0.03in}}
\hline
\noalign{\vspace{0.03in}}
JS91  & $10.0 \pm 20\%$ **)& ... & $2.23 \pm 0.2$ & $4250 \pm 5\%$ *) &  
$27 \pm 20\%$ & $1.0 \pm 20\%$ &  $2.0 \cdot 10^{-10}$ \\

GLG99 & $11.3 \pm 2\%$ **)& ... & $2.22 \pm 0.02$ & $4290 \pm 1\%$ &  
$23 \pm 2\%$ *) & $1.0 \pm 20\%$ & ... \\

DVW03 & $11.3 \pm 2\%$ ***)& ... & $2.29 \pm 0.07$ & $4320 \pm 3\%$ *) &
$25 \pm 6\%$  & $1.2 \pm 50\%$ & ... \\

adopted: & $11.3 \pm 2\%$ & $0.55 \pm 0.05$ & $2.23 \pm 0.02$ & $4290 \pm 1\%$ &  
$23 \pm 6\%$ *)  & $1.10 \pm 6\%$  & $2.5 \cdot 10^{-10}$ \\
\noalign{\vspace{0.03in}}
\hline
\end{tabular}
\end{center} 
\vspace{-0.1cm}
\noindent
Notes: *) derived value within the set $L_*$, $T_{\rm eff}$, $R_*$ (see text),
      **) using spectrophotometry,
     ***) using IR photometry.
     A colon (:), if given, indicates a value with an unknown error bar, which
     in some cases may be relatively large.
  \end{table*}

\section{Individual case studies}

\subsection{$\alpha$~Boo (K1.5~III)}

The star $\alpha$~Boo has long been known as a mildly metal-poor giant  
with [Fe/H] = -0.50  (Decin et al. 2003) [DVW03].  Spectrophotometry yields
$T_{\rm eff} = 4290$~K (Griffin \& Lynas-Gray 1999) [GLG99], which is accurate
within approximately 50~K. As previously suggested by Charbonnel et al. 
(1998), Arcturus is probably an old, low-mass RGB giant:
the metal-poor RGB coincides with the AGB of solar abundance in $T_{\rm eff}$,
whereas a low-mass star spends a significant time on the RGB, making this
solution by far the most likely one.  Furthermore, $\alpha$~Boo's proximity
allows a precise distance measurement (by Hipparcos) of $d = 11.3$~pc and
permits very accurate physical parameter determinations.  All this makes
$\alpha$~Boo a very interesting case; however, it differs from the other
objects considered that its observed mass-loss rate has
not been derived from a spatial resolution study.  Hence,  
it remains more uncertain.

With $V = -0.04$, a distance modulus of 0.26, and $BC = 0.55$, 
Arcturus has a luminosity of $170~L_{\odot}$, or $M_{\rm bol} = -0.85$, 
which yields a flux-related radius of $23~R_{\odot}$, consistent with 
that given by Griffin \& Lynas-Gray (1999), and corresponding to an angular 
diameter of $\theta = 18.9$ mas (see Table 1).
This is in very good agreement with the LBI (Long Baseline Interferometry)
uniform disk diameter (hereafter referred to as UDD), as
in visual light, UDDs of $19.0 \pm 0.2$ mas and 
$21.0 \pm 0.3$ mas have been measured (Richichi \& Percheron 2002). 
Note that we compare UDDs and flux radii in a consistent manner 
for all stars for reasons of simplicity. 
In fact, giant and supergiant radii are ill-defined quantities
which are often critically dependent on how they were derived.
This becomes particularly obvious for $\alpha$~Ori (see Sect. 3.6). 

As a RGB giant, Arcturus' luminosity and its effective temperature is fitted
best by an evolution track for $1.13~M_{\odot}$, using a reduced metallicity of
$Z = 0.01$ (see Fig.~1).  RGB stars of masses within $\pm 0.05~M_{\odot}$ of
that value would still be within an uncertainty of $T_{\rm eff}$ of 1\%.
Considering the possible miss-match between the grid metallicity used and the
observed [Fe/H] value for Arcturus --- slightly lower and close to
$Z = 0.007$ --- we arrive at a slightly lower mass of $1.10~M_{\odot}$, with a
total error of $\pm 0.06~M_{\odot}$.  Based on these values, we obtain a
surface gravity of $\log{g} = 1.76 \pm 0.05$ (cgs); see Table 7 for a summary
of the adopted parameters.

For the above set of parameters, the new relation by Schr\"oder \& Cuntz
(2005) with a calibrated value of $\eta = 0.8 \cdot 10^{-13}$ suggests
a mass-loss rate of $4.0 \cdot 10^{-10}$ {\Myr}.  This is well within the
uncertainty range of the observed mass-loss rate, while the old ``Reimers law"
(with a re-calibrated value of $\etaR = 2 \cdot 10^{-13}$) as well as all
other relations but one (see Table 8) result in much larger discrepancies.

As for the observed mass-loss rate, radio continuum emission measures
at 6~cm and 2~cm wavelength yielded ionized mass-loss rates between
$6.9 \cdot 10^{-11}$ and $8.4 \cdot 10^{-11}$ {\Myr}
(Drake \& Linsky 1983, 1986).  As far as the total mass-loss 
rate is concerned, we would need to know the average ionization ratio
of the CS envelope.  However, no
sophisticated model exists for the CS matter, where
the radio emission originates. But we may speculate that, due to
radiative cooling and NLTE processes, the ionization fraction decreases
considerably from close to 100\% in the mid chromosphere to low values
further out in the wind (Ayres \& Linsky 1975; Cuntz 1990). Indeed, Ayres
et al. (2003) presented observational evidence that the circumstellar 
matter of Arcturus is acting as a ``cool absorber" to the radiative 
emission of the hot 
chromospheric gas buried beneath.  This would imply a smaller ionization
degree further outward, indicating a larger total mass-loss rate
than the ionized mass loss obtained from the above radio observations.

In their study of a number of cool giants with mass loss, 
Judge \& Stencel (1991) [JS91] adopted a value of 
$2.0 \cdot 10^{-10}$ {\Myr} for the total mass loss of
Arcturus, as originally derived by Drake (1985) from Mg~II emission line 
profile modelling and in line with the aforesaid. Correcting their 
distance value of 10.0~pc to the actual distance of 11.3~pc
translates into a mass-loss rate of $2.5 \cdot 10^{-10}$ {\Myr} --- noting
that these measures are based on areal (angular) emission and, hence, the 
derived mass-loss rates scale with the distance squared.
    
While Mg~II is probably the dominant ionization stage of Mg in most of the
CS envelope, it is well-known that there are large ambiguities in the line 
modelling process.  For example, different choices of turbulent velocity
can all well result in good matches of the observed line profile, however,
nonetheless corresponding to different mass-loss rates.  Hence, we consider
the above discussed rate to be uncertain by at least a factor of 2.

  \begin{figure}
  \vspace{7cm}
  \includegraphics{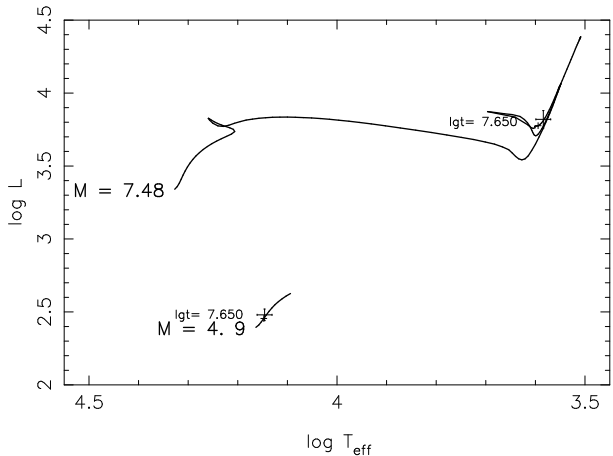}
  \caption{The observed HRD positions of both components of
  the $\zeta$~Aur-type system 32 Cygni are  
  matched very well by evolution tracks of 4.13 $M_{\odot}$
  and 7.58 $M_{\odot}$ (initial) mass, at an age of $4.4 \cdot 10^7$~yr,
  using an enhanced metallicity of $Z = 0.03$.}
  \end{figure}

\begin{table*}[bht]
  \caption{Recently published physical parameters of 32~Cyg.}

  \begin{center}
    \leavevmode
    \footnotesize
    \begin{tabular}{l c c c c c c c}
\hline \hline
\noalign{\vspace{0.03in}}
Source  & $d$ & $BC+A_v$ & $\log{L_*/L_{\odot}}$ & $T_{\rm eff}$ & $R_*$ & 
     $M_*$ & ${\dot{M}_{\rm obs}}$ \\
\omit  & [pc]& \omit & \omit & [K] & [$R_{\odot}$] & 
         [$M_{\odot}$]  & [{\Myr}] \\
\noalign{\vspace{0.03in}}
\hline
\noalign{\vspace{0.03in}}
CHR83 & $383 \pm 10\%$ & 1.0 & $3.82\pm0.2$ *)& $3800 \pm 5\%$ &  
$188 \pm 10\%$ & $8.0 \pm 15\%$ & $2.8 \cdot 10^{-8}$ \\

B90, B01 & ...  & 1.0 & $3.82\pm0.02$ *)& $3800 \pm 5\%$ &  
$188 \pm 10\%$ & $8.0 \pm 15\%$ & $1.3 \cdot 10^{-8}$ \\

adopted: & $360 \pm 5\%$ & $1.05 \pm 0.07$ & $3.82\pm0.08$ *)& $3840 \pm 3\%$ &
$184 \pm 6\%$ & $7.45 \pm 4\%$ & $1.3 \cdot 10^{-8}$ \\
\noalign{\vspace{0.03in}}
\hline
\end{tabular}
  \end{center}
\vspace{-0.1cm}
\noindent
  Notes: See Table 1.
  \end{table*}

\subsection{32 Cyg (K5 Ib)}

The wind of 32 Cyg, a K5 supergiant belonging to a spectroscopic binary
system, has been well studied by means of the hot main sequence companion 
(Baade 1998, 1990a) by using IUE and HST spectra. The flux of the secondary 
dominates the UV spectrum of the binary and, hence, it acts as an orbiting 
light source, allowing to probe the circumstellar matter in the line of sight. 
Compared to the very similar $\zeta$ Aur and 31 Cyg supergiants, the 32 Cyg 
system has the least hot B star companion. Therefore, ionization effects by 
its radiation are less severe and less likely to affect any mass-loss rates
derived from observation.

A mass-loss rate of 32 Cyg of $1.5 \cdot 10^{-8}$ {\Myr} has been 
obtained by Baade (1990a, b) from high-resolution IUE spectra by 
modelling the resonance line scattering of singly ionized metals. 
Baade et al. (2001) revised this value to $1.3 \cdot 10^{-8}$ {\Myr}
while considering high-resolution HST spectra. 
The residual uncertainty of this rate is
mainly caused by considerable intrinsic variability and chromospheric 
density fluctuations (see Wright 1970, Schr\"oder 1983, Baade et al.~2001).
We estimate it to be within a factor of about 1.6.

The eclipse geometry of 32 Cyg presents the peculiar complication that
the secondary's projected path grazes the limb of the giant.  Hence,
the eclipse geometry does not yield the radii, but it gives the 
inclination of the system for given mass ratio and giant radius. 
In order to determine the physical parameters, we can also consider 
(1) the mass function of the giant's radial velocity curve, 
0.301 (Wright 1970), which constrains the masses and their mass ratio $q$, 
(2) the semi-major axis $a_1 \sin{i}$ of the primary orbit of 
$2.55 \cdot 10^8$~km (Wright 1970), and (3) the effective temperature
of the secondary of 14,000~K, well-determined by UV spectroscopy, and its
angular diameter of 0.077 mas (Erhorn 1990, taking extinction into account),
as well as (4) the effective temperature of 3840~K of the primary. 
The latter value was found by Levesque et al. (2005) [LMO05] from 
a comparison of the G-Band
with MARCS stellar atmospheres.  Table 2 compares the physical parameters
and mass-loss rates adopted for the giant from earlier publications, i.e., 
Che et al. (1983) [CHR83], Baade (1990a, 1990b) [B90], and
Baade et al. (2001) [B01], with this paper.

Hipparcos measured a parallax of $\pi = 2.94 \pm 0.6$ mas.  It puts 32~Cyg 
at a distance of 340 (+90/-60) pc, smaller than the 383~pc suggested by 
CHR83.  Since large relative parallax errors are asymmetric in distance, 
measured distances appear, on average, a little smaller 
(``Malmquist bias"), see Malmquist (1936).  Hence, we find $d = 360$~pc 
a more likely value, from which for the secondary a radius of
$3.0 R_{\odot}$ and a luminosity of $302~L_{\odot}$ is derived.
For the primary ($V = 4.0$), we adopt $L_* = 6600~L_{\odot}$, which results 
from a radius of 184~$R_{\odot}$ and from $T_{\rm eff} = 3840$~K, 
and which is consistent with $BC+A_v = 1.05$.  The corresponding 
$\theta = 4.75$ mas is then in excellent agreement with the UDD, measured
by LBI, of $4.8 \pm 0.1$~mas listed by Richichi \& Percheron (2002).

Next, we review the system parameters of 32 Cyg by means of evolution models,
which are constructed to match the HRD positions of both stars for their
well enhanced metallicity (Taylor 1999), and taking into account the
equal age of both components.  At the same time, eclipse geometry
and mass function must be satisfied as well.  A detailed
description of this method has been presented by Schr\"oder et al.~(1997). 
We find that the secondary is fitted best by 
an evolution track with a mass of $4.13~M_{\odot}$, while the primary 
requires an initial mass of around $7.58~M_{\odot}$ (as shown in Fig.~2). 
This mass is reduced by mass loss, as given by the relation
of Schr\"oder \& Cuntz (2005), to $7.46~M_{\odot}$ by the time the star
has reached about the end of its core helium burning, i.e., after
$4.4 \cdot 10^7$~yr of age.  These mass values reproduce the mass function
and the eclipse geometry very well with $q = 1.80$ and $i = 80^{\circ}$.
This good multiple match not only yields the masses of the 32~Cyg components
with unprecedented accuracy, it also reduces considerably the uncertainties
of luminosity, radius and $T_{\rm eff}$ (see Table 2 for estimated errors).

With a mass of $7.45~M_{\odot}$ and a radius of $184~R_{\odot}$, 
the surface gravity of the 32~Cyg giant is $\log{g} = 0.78$ (cgs). 
The predicted mass-loss rates, as given by the relation
of Schr\"oder \& Cuntz (2005) and by various 
other formulas (see Sect. 2.1), are compared in Table 8. 
The law by Schr\"oder \& Cuntz (2005) suggests
a mass-loss rate of $2.3 \cdot 10^{-8}$ {\Myr}, which is consistent
with the observed value.
The original ``Reimers law" (with $\etaR = 2 \cdot 10^{-13}$), however,
suggests $3.2 \cdot 10^{-8}$ {\Myr}, which is more than a factor of 2
above the observed value.  The predictions by the other mass-loss relations
disagree by much larger factors (see Table 8), and are well outside the
combined observational and theoretical uncertainty bars.

  \begin{figure}
  \vspace{7cm}
  \includegraphics{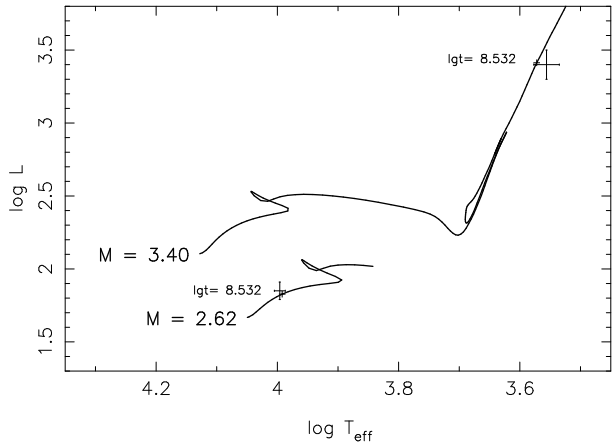}
  \caption{Observed HRD positions of the components of $\delta$~Sge and 
  evolution tracks of 2.62 $M_{\odot}$ and 3.40 $M_{\odot}$ (initial) 
  mass, using a metallicity of $Z = 0.02$.}
  \end{figure}

\begin{table*}[bht]
  \caption{Recently published physical parameters of $\delta$~Sge.}

  \begin{center}
    \leavevmode
    \footnotesize
    \begin{tabular}{l c c c c c c c}
\hline \hline
\noalign{\vspace{0.03in}}
Source  & $d$ & $BC+A_v$ & $\log{L_*/L_{\odot}}$ & $T_{\rm eff}$ & $R_*$ & 
     $M_*$ & ${\dot{M}_{\rm obs}}$ \\
\omit  & [pc]& \omit & \omit & [K] & [$R_{\odot}$] & 
         [$M_{\odot}$]  & [{\Myr}] \\
\noalign{\vspace{0.03in}}
\hline
\noalign{\vspace{0.03in}}
RS83  & $170 \pm 25\%$ & 1.62 & $3.43 \pm 0.2$ *)& $3600 \pm 5\%$ &  
$140 \pm 25\%$ & 8: & $2.0 \cdot 10^{-8}$ \\

SPE97 & $182 \pm 25\%$ & 1.81 & $3.57 \pm 0.2$ *)& $3600 \pm 5\%$ &  
$157 \pm 25\%$ & $3.4 \pm 10\%$  &  ... \\
adopted: & $150 \pm 10\%$ & $2.0 \pm 0.2$ & $3.40\pm0.10$ *)& $3600 \pm 8\%$ & 
$129 \pm 10\%$ & $3.35 \pm 10\% $ & $1.4 \cdot 10^{-8}$ \\
\noalign{\vspace{0.03in}}
\hline
\end{tabular}
  \end{center}
\vspace{-0.1cm}
\noindent
  Notes: See Table 1.
  \end{table*}

\subsection{$\delta$ Sge (M2 II)}

The star $\delta$ Sge~A is a moderate mass AGB giant, and it is a member of
a non-eclipsing spectroscopic binary system.  Spectro-photometry has been 
carried out for the A0-type companion.  The radial velocity measurements 
for both components imply an observed mass ratio of about 1:1.27 
(Griffin, private communication).  In addition, separation measurements by
speckle interferometry exist for various phases indicating an inclination of
$i = 40^\circ$ and a distance of about 180~pc.  
A study for $\delta$ Sge~A
based on these observations and on matching evolution tracks
(Schr\"oder et al.~1997) [SPE97] suggested
a luminosity of $\log{L_*/L_{\odot}} = 3.57$ (with $V = 3.95$ and $BC = 1.75$),
an effective temperature of $T_{\rm eff} = 3600$~K, a radius of
$R_* = 157~R_{\odot}$, and a mass of $M_* = 3.4~M_{\odot}$.

However, the Hipparcos parallax of $\delta$~Sge suggests a distance 
of 137 (+18/-14)~pc.  Hence, we adopt a distance of 150~pc 
($m - M = 5.73$), which is fully consistent with a B star radius of
2.85~$M_{\odot}$, considering a spectro-photometrically determined
temperature of $T_{\rm eff} = 9900$~K and angular diameter of 0.178 mas
of the companion, Erhorn 1990).  Therefore, we adopt the same $T_{\rm eff}$
of 3600~K for the $\delta$~Sge giant as used in that study.
Furthermore, with $BC = 1.75$ and an estimated extinction (IS and CS 
combined) of $A_v = 0.25$, we derive a primary luminosity of 
$\log{L_*/L_{\odot}} = 3.40$ (or $M_{\rm bol}$ = -3.68) and a 
(flux-) radius of $R_* = 129~R_{\odot}$.
This yields an angular diameter of $\theta = 7.8$ mas, in excellent agreement
with the LBI measurement of best accuracy for the UDD ($7.8 \pm 0.3$ mas) 
listed by Richichi \& Percheron (2002).  

We estimate the remaining uncertainties to be 10\% for the geometry 
($R_*$, $d$) and mass, 8\% for $T_{\rm eff}$, and 20\% for $L_*$. 
Within these uncertainties, the above physical parameters
are well matched by evolution tracks (for $Z = 0.02$ and accounting for 
mass loss) for todays component masses of about 2.62 and 
$3.35~M_{\odot}$ (see Fig.~3).  The relatively low luminosity of the M2
primary alone would rather suggest a lower mass of $\approx 3 M_{\odot}$,
but this value is incompatible with the finding that the
secondary component is not sufficiently advanced in its 
main-sequence evolution.  Hence, we thus arrive at a surface gravity of
$\log{g} = 0.74 \pm 0.10$ (cgs).

The mass loss of $\delta$~Sge has been observed by means of wind lines 
in the UV spectrum of the companion probing the wind at different 
phases without any significant ionization by its radiation.
Reimers \& Schr\"oder (1983) [RS83] found rates of $1.8 \cdot 10^{-8}$ {\Myr}
for a phase close to conjunction (July 31, 1980), and of 
$2.5 \cdot 10^{-8}$ {\Myr} (March 30, 1982) using
IUE high resolution spectra.  Back then, Reimers \& Schr\"oder assumed a
much more inclined geometry (with $i=70^\circ$), implying a line-of-sight 
which passes too close by the primary in the July '80 phase. 
With an approximate correction for an inclination of $i=40^\circ$, 
both quoted phases imply a mass-loss rate of about   
$2.5 \cdot 10^{-8}$ {\Myr}.

But, in addition, Reimers \& Schr\"oder assumed a larger mass ratio of
$\mu \approx 3$. 
Since the relative orbit size $a$ is scaled with $(1+\mu)$, and since 
the wind-line interpretation mainly depends on the absorbing column 
density, the derived densities ($n$) scale, roughly, as  
$(1+\mu)^{-1}$ and the mass-loss rates $(\propto n \cdot v_w \cdot a^2)$ 
as $(1+\mu)$, with $v_w$ as wind speed.
Hence, for a much smaller mass-ratio of $\mu = 1.27$, 
we finally arrive at an updated mass loss of 
$1.4 \cdot 10^{-8}$ {\Myr} (see Table 3).  We estimate this
rate to be uncertain by a factor of 2, mainly due to intrinsic 
variability of the outflow (e.g., Reimers \& Schr\"oder 1983)
and remaining uncertainties in population ratios due to ionization 
and excitation processes.

For the above parameters, the mass-loss relation by Schr\"oder \& Cuntz
(2005) yields a rate of $1.15 \cdot 10^{-8}$ {\Myr}, lower than the 
observational value but well within the range of uncertainty, while the 
old ``Reimers law" (with $\etaR = 2 \cdot 10^{-13}$) suggests a higher rate of
$1.9 \cdot 10^{-8}$ {\Myr}.  See Table 8 for the other rates,
which are all significantly larger.

\begin{table*}[bht]
  \caption{Recently published physical parameters of $\alpha$~Sco.}

  \begin{center}
    \leavevmode
    \footnotesize
    \begin{tabular}{l c c c c c c c}
\hline \hline
\noalign{\vspace{0.03in}}
Source  & $d$ & $BC+A_v$ & $\log{L_*/L_{\odot}}$ & $T_{\rm eff}$ & $R_*$ & 
     $M_*$ & ${\dot{M}_{\rm obs}}$ \\
\omit  & [pc]& \omit & \omit & [K] & [$R_{\odot}$] & 
         [$M_{\odot}$]  & [{\Myr}] \\
\noalign{\vspace{0.03in}}
\hline
\noalign{\vspace{0.03in}}
KR78  & $180 \pm 15\%$ & 2.0 & 4.68 & $3550 \pm 10\%$ &  
$575 \pm 20\%$ *)& 18: & $0.7 \cdot 10^{-6}$ \\

HN83  & $180 \pm 15\%$ & ... & ... & ... &
... & ... & $2.0 \cdot 10^{-6}$ \\

HHR87 & $180 \pm 15\%$ & 2.0 & 4.68 & $3550 \pm 10\%$ &  
$575 \pm 20\%$ *)& 18: & $1.0 \cdot 10^{-6}$ \\

adopted: & $180 \pm 15\%$ & $2.0 \pm 0.2$ & $4.76 \pm 0.12$ &
 $3400 \pm 6\%$ &  
 $703 \pm 15\%$ *)& $12 \pm 20\%$ & $1.5 \cdot 10^{-6}$ \\
\noalign{\vspace{0.03in}}
\hline
\end{tabular}
  \end{center}
\vspace{-0.1cm}
\noindent
  Notes: See Table 1.
  \end{table*}

\subsection{$\alpha$ Sco (M1.5 Iab)}

The star $\alpha$~Sco~A, Antares, is very similar to Betelgeuse (see Sect. 3.6)
in mass, luminosity and spectral type --- only slightly hotter and, presumably,
a bit more massive.  Its distance is well known (180~pc, $\pi = 5.4 \pm 
1.7$~mas) and its circumstellar envelope is probed by a hot (18,500~K) but
distant (2.9" angular separation) companion on the near side of its orbit.
Kudritzki \& Reimers (1978) [KR78] used optical wind absorption lines in
the spectrum of the companion and simple ionization considerations to derive
a mass-loss rate of $7 \cdot 10^{-7}$ {\Myr} for $\alpha$~Sco~A.  Due to the
wide orbit of the binary system, radiative ionization by the B2.5~V
companion concerns only a small part of the cool wind.  Hence, this star
can accurately be modelled, including the height-dependent ionization degree
in the line of sight (van der Hucht et al. 1980).

Hagen et al. (1987) [HHR87] used IUE spectra and redetermined the mass-loss 
rate from carefully selected, not blended circumstellar absorption lines in
the spectrum of $\alpha$~Sco~B.  They found a value
of about $1.0 \cdot 10^{-6}$ {\Myr}, with a remaining uncertainty 
of a factor of 2.  From spacially resolved radio observations with VLA,
a very different approach, Hjellming \& Newell (1983) [HN83] deduced 
a mass loss rate of about $2.0 \cdot 10^{-6}$ {\Myr} with a similar 
uncertainty.  Hence, we adopt an observed mass-loss rate for $\alpha$~Sco~A
of about $1.5 \cdot 10^{-6}$ {\Myr} (see Table 4).

The physical parameters of the supergiant $\alpha$~Sco~A are fairly 
well constrained.  Here we adopt $V \approx 1.02$ (variable), $m-M = 6.28$, a 
combined $BC$ and $A_v$ of 2.0 (similar to 2.1 for $\alpha$~Ori, see below), 
resulting in $M_{\rm bol} = -7.2$ and $\log{L_*/L_{\odot}} = 4.76 (\pm 0.12)$.
Spectral type and $V-K$ index are only slightly earlier (smaller) than 
those of $\alpha$~Ori, indicating that its effective temperature should be
slightly higher.  Compared to the purely empirical, UDD-based 
$\alpha$~Ori Model~A (see below), for $\alpha$~Sco~A we should expect a
$T_{\rm eff} \simeq 3400$~K ($\pm 200$~K), giving a flux radius of 
$R_* = 703~R_{\odot}$, with an estimated error of 15\%, and 
$\theta = 36.4$ mas. This value is in good agreement with the 
LBI-measured UDDs of $\alpha$~Sco~A, 
which range from 29 to 45 mas (Richichi \& Percheron 2002) 
and with the $V-K$-related value of 37.7 mas by the same authors.

For a mass of approximately $12~M_{\odot}$, as indicated by our best-matching
evolution model, the surface gravity of $\alpha$~Sco~A
is $\log{g} = -0.18 \pm 0.11$ (cgs).  With these values, the mass-loss relation
by Schr\"oder \& Cuntz (2005) yields a rate of $1.5 \cdot 10^{-6}$ {\Myr},
in perfect agreement with the observed value, while the old ``Reimers law"
(with $\etaR = 2 \cdot 10^{-13}$) predicts $6.6 \cdot 10^{-7}$ {\Myr}, a rate
that is much too small (see Table 8 for the results from other relations).

\begin{table*}[bht]
  \caption{Recently published physical parameters of $\alpha^1$~Her.}

  \begin{center}
    \leavevmode
    \footnotesize
    \begin{tabular}{l c c c c c c c}
\hline \hline
\noalign{\vspace{0.03in}}
Source  & $d$ & $BC+A_v$ & $\log{L_*/L_{\odot}}$ & $T_{\rm eff}$ & $R_*$ & 
     $M_*$ & ${\dot{M}_{\rm obs}}$ \\
\omit  & [pc]& \omit & \omit & [K] & [$R_{\odot}$] & 
         [$M_{\odot}$]  & [{\Myr}] \\
\noalign{\vspace{0.03in}}
\hline
\noalign{\vspace{0.03in}}
R77     & 59: & 3.2 & ... & $2880 \pm 10\%$ *)&  
$140 \pm 25\%$ & 8: & $1.0 \cdot 10^{-7}$ \\

TR93    & 70: & 3.2 & ... & $2880 \pm 10\%$ *)&  
$157 \pm 25\%$ & $3.4 \pm 10\%$ & $1.5 \cdot 10^{-7}$ \\

LMO05   & ($120 \pm15\%$) & 3.9: & 4.33 & 3450: &
410: *) & ... & ... \\ 

adopted: & $120 \pm 15\%$ & $2.9 \pm 0.4$   & $3.92 \pm 0.15$ & $2800 \pm 10\%$ *) &
$387 \pm 20\%$ & $2.15 \pm 10\%$ & $3.0 \cdot 10^{-7}$ \\
\noalign{\vspace{0.03in}}
\hline
\end{tabular}
  \end{center}
\vspace{-0.1cm}
\noindent
  Notes: See Table 1.  $L_*$ and $R_*$ of [LMO05] are based on  $d = 120$~pc,
according to the Hipparcos parallax.
  \end{table*}

\subsection{$\alpha^1$ Her (M5 Ib)}

The wind of $\alpha^1$~Her, a M-type supergiant, is just below the critical
luminosity for becoming dust-driven (see Schr\"oder et al. 1999). 
Its physical structure is probed by a spatially well resolved companion star,
$\alpha^2$~Her, separated by 4.7" from its primary. Reimers (1977) [R77] 
adopted a distance of 59~pc and a mass-loss rate of 
$\dot{M} = 1.1 \cdot 10^{-7}$ {\Myr} from the wind absorption lines. 
These are seen in the composite spectrum of $\alpha^2$~Her, itself 
a spectroscopic binary (i.e., $\alpha^2$~Her A \& B).

A detailed study was provided by Thiering \& Reimers (1993) [TR93].  They used
IUE spectra for deriving the component's brightnesses and effective 
temperatures. Their mass-loss rate of the M supergiant is based on the 
line-of-sight line absorption by using mass column densities and the 
wind velocity structure, also considering ionization effects. For a distance
of $d = 70$~pc, they found $\dot{M} = 1.5 \cdot 10^{-7}$ {\Myr}
with a 20\% uncertainty (see Table 5).
Mauron \& Caux (1992) used observations of spatially 
resolved K~I and Na~I line scattering to derive an update value of 
$\dot{M} = 1.0 \cdot 10^{-7}$ {\Myr}, assuming a distance of 
60~pc.  Since all historic distance values of {$\alpha^1$~Her seem 
to be gross underestimates, according to modern measurements 
(see below), a significant adjustment of these observed mass-loss 
rates is required.

  \begin{figure}
  \vspace{7cm}
  \includegraphics{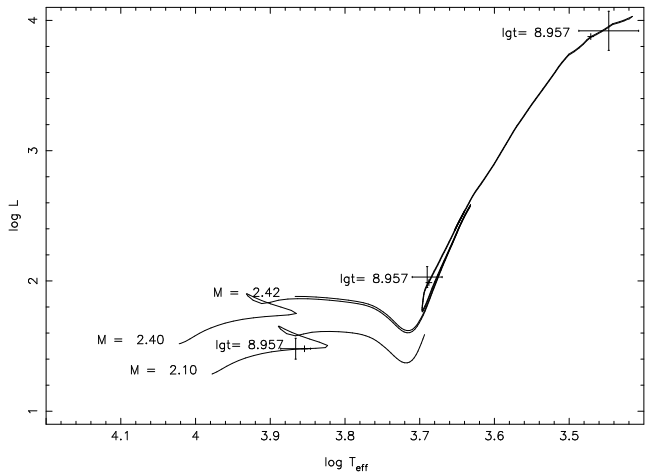}
  \caption{Observed HRD positions of the three components of $\alpha$~Her and 
  evolution tracks of 2.10, 2.40 and 2.42 $M_{\odot}$ (initial) 
  mass, using a metallicity of $Z = 0.02$.}
  \end{figure}

\begin{table*}[bht]
  \caption{Recently published physical parameters of $\alpha$~Ori.}

  \begin{center}
    \leavevmode
    \footnotesize
    \begin{tabular}{l c c c c c c c}
\hline \hline
\noalign{\vspace{0.03in}}
Source  & $d$ & $BC+A_v$ & $\log{L_*/L_{\odot}}$ & $T_{\rm eff}$ & $R_*$ & 
     $M_*$ & ${\dot{M}_{\rm obs}}$ \\
\omit  & [pc]& \omit & \omit & [K] & [$R_{\odot}$] & 
         [$M_{\odot}$]  & [{\Myr}] \\
\noalign{\vspace{0.03in}}
\hline
\noalign{\vspace{0.03in}}
JS91    & 200: & **) & 5.0: & 3900: *) &  
   860: & 10: & $4 \cdot 10^{-6}$: \\

HBL01   & $131 \pm 20\%$ & **) & $4.73 \pm 0.15$ & 3140: *)&  
$789 \pm 20\%$ & ... & $3.1 \cdot 10^{-6} \pm 40\%$ \\

LMO05   & ($131 \pm 20\%$) & $2.2 \pm 0.3$ & $4.81 \pm 0.15$ & 3650: &
637: *) & ... & ... \\

Model~A & $131 \pm 20\%$ & $2.1 \pm 0.3$ & $4.73 \pm 0.15$ &$3340\pm6\%$ *)& 
$695 \pm 20\%$ &  $10 \pm 20\%$ & $3.1 \cdot 10^{-6} \pm 40\%$ \\

Model~B & $131 \pm 20\%$ & $2.2 \pm 0.3$ & $4.81 \pm 0.15$ & $3650\pm10\%$ &
$637 \pm 20\%$ *)&  $10 \pm 20\%$ & $3.1 \cdot 10^{-6} \pm 40\%$ \\
\noalign{\vspace{0.03in}}
\hline
\end{tabular}
  \end{center}
\noindent
  Notes: See Table 1.  $L_*$ and $R_*$ of [LMO05] are based on $d = 131$~pc,
according to the Hipparcos parallax.
  \end{table*}

\vspace{1cm}

\begin{table*}[bht]
  \caption{Adopted physical parameters for the various stars.}
  \label{tab:table}
  \begin{center}
    \leavevmode
    \footnotesize
    \begin{tabular}{l c c c c c c}
\hline \hline
\noalign{\vspace{0.03in}}
Object  & $d$ & $\log{L_*/L_{\odot}}$ & $T_{\rm eff}$ & $R_*$ & 
                                                  $M$ & $\log{g_*}$ \\
\omit  & [pc] & \omit & [K] & [$R_{\odot}$] & [$M_{\odot}$] & \omit \\ 
\noalign{\vspace{0.03in}}
\hline
\noalign{\vspace{0.03in}}
$\alpha$~Boo      & $11.3 \pm 2\%$  & $2.23 \pm 0.02$ & $4290  \pm 1\%$
                  &  $23  \pm 6\%$  & $1.10 \pm 6\%$  & $1.76  \pm 0.05$ \\
32~Cyg            & $360  \pm 5\%$  & $3.82 \pm 0.08$ & $3840  \pm 3\%$
                  & $184  \pm 6\%$  & $7.45 \pm 4\%$  & $0.78  \pm 0.06$ \\
$\delta$~Sge      & $150  \pm 10\%$ & $3.40 \pm 0.10$ & $3600  \pm 8\%$
                  & $129  \pm 10\%$ & $3.35 \pm 10\%$ & $0.74  \pm 0.10$ \\
$\alpha$~Sco      & $180  \pm 15\%$ & $4.76 \pm 0.12$ & $3400  \pm 6\%$
                  & $703  \pm 15\%$ &   $12 \pm 20\%$ & $-0.18 \pm 0.11$ \\
$\alpha^1$~Her    & $120  \pm 15\%$ & $3.92 \pm 0.15$ & $2800  \pm 10\%$
                  & $387  \pm 20\%$ & $2.15 \pm 10\%$ & $-0.41 \pm 0.19$ \\
$\alpha$~Ori~[A]  & $131  \pm 20\%$ & $4.73 \pm 0.15$ & $3340  \pm 6\%$
                  & $695  \pm 20\%$ &   $10 \pm 20\%$ & $-0.25 \pm 0.23$ \\
$\alpha$~Ori~[B]  & $131  \pm 20\%$ & $4.81 \pm 0.15$ & $3650  \pm 10\%$
                  & $637  \pm 20\%$ &   $10 \pm 20\%$ & $-0.17 \pm 0.14$ \\
\noalign{\vspace{0.03in}}
\hline
\end{tabular}
  \end{center}
\vspace{-0.1cm}
\noindent
  Notes: The estimated relative uncertainties are given in \% or dex.
Relationships between the uncertainties are considered, if applicable.
For instance, if the distance $d$ is uncertain, the radius $R_*$ scales as
$\propto d$ and the luminosity $L_*$ as $\propto d^2$.
  \end{table*}

\begin{table*}[bht]
  \caption{Observed versus predicted mass-loss rates in log({\Myr}).}
  \label{tab:table}
  \begin{center}
    \leavevmode
    \footnotesize
    \begin{tabular}{l c c c c c c c}
\hline \hline
\noalign{\vspace{0.03in}}
Reference & $\alpha$~Boo      & 32~Cyg            & $\delta$~Sge    & $\alpha$~Sco 
          & $\alpha^1~$Her    & $\alpha$~Ori~[A]& $\alpha$~Ori~[B]\\
\noalign{\vspace{0.03in}}
\hline
\noalign{\vspace{0.03in}}
R75       & $-9.14 \pm 0.03$ & $-7.49 \pm 0.07$ & $-7.71 \pm 0.14$  & 
$-6.18 \pm 0.13$ & $-6.52 \pm 0.14$ & $-6.13 \pm 0.14$ & $-6.09 \pm0.17$ \\
L81       & $-8.83 \pm 0.03$ & $-6.85 \pm 0.10$ & $-7.19 \pm 0.20$  & 
$-5.37 \pm 0.16$ & $-5.97 \pm 0.20$ & $-5.33 \pm 0.19$ & $-5.24 \pm0.21$ \\
dJNH88    &   ...            & $-7.15 \pm 0.11$ & $-7.53 \pm 0.33$  & 
$-6.05 \pm 0.11$ & $-6.27 \pm 0.36$ & $-6.09 \pm 0.19$ & $-5.95 \pm0.17$ \\
NdJ90     & $-9.73 \pm 0.03$ & $-6.62 \pm 0.10$ & $-7.40 \pm 0.20$  & 
$-4.78 \pm 0.18$ & $-6.30 \pm 0.21$ & $-4.84 \pm 0.21$ & $-4.76 \pm0.23$ \\
SC05      & $-9.38 \pm 0.03$ & $-7.63 \pm 0.10$ & $-7.94 \pm 0.20$  & 
$-5.81 \pm 0.25$ & $-6.23 \pm 0.29$ & $-5.71 \pm 0.29$ & $-5.60 \pm0.34$ \\
\noalign{\vspace{0.03in}}
\hline
\noalign{\vspace{0.03in}}
Observed & $-9.60 \pm 0.3$ & $-7.82 \pm 0.2$ & $-7.85 \pm 0.3$ &
$-5.82 \pm 0.3$ & $-6.50 \pm 0.3$ & $-5.51 \pm 0.2$ & $-5.51 \pm 0.2$ \\
\noalign{\vspace{0.03in}}
\hline
\end{tabular}
  \end{center}
\noindent
  Notes: The estimated uncertainties of the observed and predicted values
are given in dex.  
For R75, $\etaR = 2 \cdot 10^{-13}$ is used (see Schr\"oder \& Cuntz 2005).
  \end{table*}

The Hipparcos parallax ($\pi = 8.5 \pm 2.8$~mas) now results in a distance of
120~pc ($\pm 30$~pc) for the $\alpha$~Her system.  In fact, this value is
perfectly consistent with a radius of $3.2 \pm 0.1 R_{\odot}$ for the A9~V 
star $\alpha^2$~Her~B with $T_{\rm eff} = 7350$~K and 
$\log{L_*/L_{\odot}} = 1.49$, as previously derived by Thiering 
\& Reimers (1993) from $V = 6.6$, $M-m = -5.4$, $A_v = 0.1$, and
$BC = 0.1$. Furthermore, these properties compare well with an evolution 
model for a $2.1 M_{\odot}$ mass star at an advanced age of about 
$9 \cdot 10^8$~yr.  These values for
the age and distance of the system are well consistent with the
evolution model for $\alpha^2$~Her~A, a G9~III giant of 
$2.40 M_{\odot}$ with $T_{\rm eff} = 4900$~K (Thiering \& Reimers 1993)
and $\log{L_*/L_{\odot}} = 2.03$ (for $V = 5.6$ and $BC = 0.46$).
These good matches
(see Fig.~4) therefore confirm the Hipparcos distance of 120~pc and,
consequently, we adopt a reduced uncertainty, i.e., about 15\%.

Recently, Levesque et al. (2005) [LMO05] derived $T_{\rm eff} = 3450$~K, 
$BC = 2.49$ and $A_v = 1.40$ for the M-type giant primary $\alpha^1$~Her 
by detailed atmospheric modelling using the strengths of the TiO bands.
The resulting physical properties are given in Table 5, assuming a distance of 
120~pc and $V = 3.2$ for the M supergiant alone (with $V = 3.06$ for 
the system of $\alpha$~Her, according to SIMBAD).
In general, the M-type $T_{\rm eff}$-scale by Levesque et al. (2005)
is less cool, and $BC$ values are less large (but accompanied by unusually
large $A_v$ values), as those of many other authors.  In the case of 
$\alpha^1$~Her, this does not result in a smaller radius, however.  It
would be $R \approx 410 R_{\odot}$, considering the unusually large absorption 
of $A_v = 1.4$ required by Levesque et al. (2005) to attain a consistent
atmospheric model.  This implies, at the same time, a much larger luminosity.

The LBI-measured angular diameter in the $V$ band (UDD; see Richichi \& 
Percheron 2002) of 30~mas yields a radius of 387 $R_{\odot}$ 
(for $d = 120$~pc).  Hence, based on this empirical angular diameter, 
we adopt the following parameters for $\alpha^1$~Her (Table 5): 
With $V = 3.2$ for the M-giant alone and a combined $BC$ and $A_v$ of 2.9, 
we find $\log{L_*/L_{\odot}} = 3.92$.
With a radius of $387~R_{\odot}$, we then derive 
$T_{\rm eff} = 2800$~K, which is in reasonable agreement with 
near tip-AGB evolution models.  The M supergiant's mass can also
be well constrained (within about 10\%),  
since $\alpha^1$~Her must have evolved from a star of slightly larger
initial mass than the G9 giant, i.e., about $2.4~M_{\odot}$.  Therefore,
we obtain a present mass of $2.15~M_{\odot}$, considering the mass loss
given by our evolution model (see Fig.~4).  Based on these values, we find
a surface gravity of $\log{g} = -0.41 \pm 0.19$ (cgs).

The aforementioned mass-loss rates derived from the line-of-sight absorption 
of wind-lines scale linearly with the system dimensions, which in this
case are proportional to the adopted values of $d$.  Hence, a distance of
$d = 120$~pc yields observed mass-loss rates of $2.2 \cdot 10^{-7}$
and $2.6 \cdot 10^{-7}$ {\Myr} for the studies of Reimers (1977) and
Thiering \& Reimers (1993), respectively.  The rate derived by
Mauron \& Caux (1992), however, scales with the distance squared, 
since it is based on emission measure, and thus translates into
$4.0 \cdot 10^{-7}$ {\Myr}.  Consequenty, we adopt
$3.0 \cdot 10^{-7}$ {\Myr}, which is considered uncertain
by a factor of 2.

With the adopted physical parameters for $\alpha^1$~Her, our new mass-loss
relation (Schr\"oder \& Cuntz 2005) results in a mass-loss rate of  
$5.9 \cdot 10^{-7}$ {\Myr}, while the old ``Reimers law"
(with $\etaR = 2 \cdot 10^{-13}$) yields $3.0 \cdot 10^{-7}$ {\Myr} 
 --- see Table 8, also for other mass-loss predictions.  In this particular 
case, it seems that the old ``Reimers law" is more accurate than our 
new relation. Nonetheless, both mass-loss predictions are consistent 
with the observed value within the uncertainty bars.  
Furthermore, if in fact $T_{\rm eff}$ were larger by 10\%, 
which is principally possible, and hence the radius were smaller by 20\%,
our new relation would result in the closer match.

\subsection{$\alpha$ Ori (M2 Iab)}

A good test candidate for the various empirical mass-loss formulas
is the well-studied massive cool supergiant $\alpha$~Ori (Betelgeuse).
Betelguese has some circumstellar dust, but it is considered to be
not dynamically important, owing to the fact that Betelgeuse's
luminosity is just below its critical luminosity value to initiate a
truly dust-driven wind.

Mass-loss rates derived from observation have improved a lot since,
e.g., Judge \& Stencel (1991) [JS91], when models based on the 
total line flux were used.  A mass-loss rate of
$3.1 (\pm 1.3) \cdot 10^{-6}$ {\Myr} has been derived by
Harper et al. (2001) [HBL01], using spatially resolved radio flux data
and assuming a distance of $d = 131$~pc ($\pm 20$\%), based on the
Hipparcos parallax.  From spectrophotometry, they also derive a corresponding
luminosity of $L_* = 5.35 \cdot 10^4 L_{\odot}$, or $M_{\rm bol} = -7.10$,
implying a combined $BC$ and $A_v$ of 2.1 ($V = 0.6$).  With an angular 
diameter (based on the purely empirical UDD) of 56 ($\pm 1$) mas, 
they find $R_* = 789 R_{\odot}$ and $T_{\rm eff} = 3140$~K.
By contrast, Levesque et al. (2005) [LMO05] find $T_{\rm eff} =
3650$~K and suggest $A_v = 0.62$.  On their $BC$-scale $\alpha$~Ori would
require about 1.60.  In Table 6, we list all relevant physical
properties, following Levesque et al. and based on a distance of 131~pc,
hereafter referred to as $\alpha$~Ori Model~B.  

However, the resulting radius of the spectroscopy-based Model~B,
637~$R_{\odot}$, is too small to be consistent with the most 
accurate and purely empirical UDD (LBI, 800 nm):  Richichi \& 
Percheron (2002) list $\alpha$~Ori's angular diameter
as $49.4 \pm 0.24$ mas, suggesting a radius of $695 R_{\odot}$ 
at the same 131~pc distance.  This purely empirical radius implies 
$T_{\rm eff} = 3340$~K, as listed in Table 6 as $\alpha$~Ori Model~A,
based on the above LBI measurement.  Since our best-matching evolution 
track suggests a mass of $M_* \simeq 10 M_{\odot}$ ($\pm 20$\%) for 
$\alpha$~Ori, the corresponding values for the surface gravity are
$\log{g} = - 0.25 \pm 0.23$ (cgs) (Model~A) and $- 0.17 \pm 0.14$ (cgs)
(Model~B).

For testing the mass-loss relations, we here proceed with both models, 
A and B, in order to properly reflect the whole range of parameters produced
by the wealth of observations on this particular star.
For Model~A, the new mass-loss relation
by Schr\"oder \& Cuntz (1995) yields $2.0 \cdot 10^{-6}$ {\Myr}, which is
consistent with the observed value of $3.1 \cdot 10^{-6}$ within the
uncertainty bars, while the ``Reimers law" (with $\etaR = 2 \cdot 10^{-13}$)
gives $7.4 \cdot 10^{-7}$ {\Myr}.  If Model~B is adopted, the mass-loss 
relation by Schr\"oder \& Cuntz yields $2.5 \cdot 10^{-6}$ {\Myr}, while 
the original ``Reimers law" yields $8.1 \cdot 10^{-7}$ {\Myr}.  In both cases, 
the new relation by Schr\"oder \& Cuntz (2005) matches the observations very
well, while the old ``Reimers law" suggests rates significantly too low.
In fact, the new relation by Schr\"oder \& Cuntz is best among all tested
relations, regardless whether Model~A or Model~B is selected (see Table 8).
The predictions by the other mass-loss relations are consistently too low
or too high, and are mostly well outside the combined observational and
theoretical uncertainties.


\section{Conclusions}

  \begin{figure}
  \vspace{8.0cm}
  \includegraphics{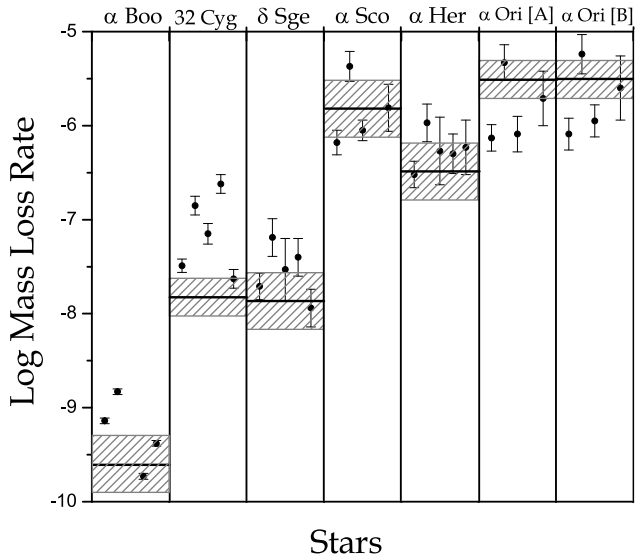}
  \caption{Comparison of observed (thick horizontal lines, uncertainties 
   indicated by dashed areas) and predicted mass-loss rates (with error bars) 
   for our target stars, as given in Table~8. The different predicted 
   mass-loss rates follow, for each star, the sequence R75, L81, dJNH88, 
   NdJ90, and SC05, from l.t.r. Note that only the values given by SC05 
   agree with the observations in {\it all} cases.
           }
  \end{figure}

It has been the aim of this paper to check the quality of our new, improved
Reimers-type mass-loss law given by Schr\"oder \& Cuntz (2005).  For this
purpose, we considered a set of well-studied giant and supergiant stars,
which are: $\alpha$~Boo (K1.5~III), 32~Cyg (K5~Ib), $\delta$~Sge (M2~II),
$\alpha$~Sco (M1.5~Iab), $\alpha^1$~Her (M5~Ib), and $\alpha$~Ori (M2~Iab).
These stars, except $\alpha$~Boo, have spatially resolved circumstellar
shells and winds --- obtained directly or by means of a companion acting as a
probing light source --- resulting in relatively accurate, empirically 
determined mass-loss rates.  In addition, they encompass a significant range
in luminosity (factor of 400), stellar radius (factor of 30), and stellar mass
(factor of 10).  As comparison, we used the empirical mass-loss relations given
by Reimers (1975), Lamers (1981), de~Jager et al. (1988) and Nieuwenhuijzen \&
de~Jager (1990).

As part of the comparison, we considered updated stellar parameters, including
stellar masses, which have been constrained using detailed stellar evolution
computations.  Further updates include recent angular diameter measurements
and stellar spectral analyses.  In addition, we considered parallax 
determinations by Hipparcos resulting in improved values for the 
stellar luminosities.  For all of our target stars, we find that the 
mass-loss rates suggested by
the relation by Schr\"oder \& Cuntz (2005) comfortably agree with the observed
values within the uncertainty bars. This is often not the case for the
other mass-loss relationships considered here (see Fig.~5).  
In many cases, those predictions are much too high or much too low, 
without any systematic trend in regard to the relationship or the target 
star. The uncertainty bars of the
empirical predictions were obtained based on a detailed error propagation
analysis also considering any mutual dependencies of variable such as
$L_*$, $R_*$, and $T_{\rm eff}$.  

Hence, we presented detailed empirical evidence in support of the
new Reimers-type mass-loss relation by Schr\"oder \& Cuntz (2005),
which we now feel safe to recommend for application to moderate,
cool (i.e., not dust-driven) stellar winds.  However, we like to 
caution that this new relation shall not be applied to predict the 
rates of stellar mass-loss physically very different from 
moderate, cool winds --- especially not to hot radiation-driven winds,
coronal mass-loss, pulsation-driven winds, and massive dust-driven winds.

Our future efforts will also include to investigate whether an underlying 
Reimers-type mass-loss mechanism could provide vital {\it initial} energy 
input for the non-pulsating dust-driven winds of the coolest AGB giants.
After all, any efficient dust-formation relies on a supplementary mechanism
needed to carry sufficient matter from the stellar photosphere to the
dust formation radius.  Such a mechanism is needed to ensure a sufficiently
high optical depth at the inner boundary of the mass-loss region, initiating
massive mass loss by radiative pressure on the dust.


\begin{acknowledgements}
This work has partially been supported by NSF grant ATM-0538278 (M.~C.).
We also acknowledge support by L. Gurdemir for his assistance with computer
graphics, and we wish to thank the anonymous referee for his helpful comments.

\end{acknowledgements}



\begin{thebibliography}{}

\bibitem[Airapetian et al.(2000)]{a00}
Airapetian, V. S., Ofman, L., Robinson, R. D., Carpenter, K., \& Davila, J.
2000, \apj, 528, 965

\bibitem[Ayres \& Linsky (1975)]{al75}
Ayres, T. R., \& Linsky, J. L. 1975, \apj, 200, 660

\bibitem[Ayres et al.(2003)]{abh03}
Ayres, T. R., Brown, A., \& Harper, G. M. 2003, \apj, 598, 610

\bibitem[Baade (1990a)]{b90a}
Baade, R. 1990a, \aap, 233, 486 

\bibitem[Baade (1990b)]{b90b}
Baade, R. 1990b, in Evolution in Astrophysics, ESA SP-310, 65 (B90)

\bibitem[Baade (1998)]{b98}
Baade, R. 1998, in Ultraviolet Astrophysics: Beyond the IUE 
          Final Archive, ESA SP-413, 325

\bibitem[Baade et al.(2001)]{betal01}
Baade, R., Kirsch, T., Reimers, D., Brown, A., Bennett, P., \&
         Harper, G. 2001, in Cool Stars, Stellar Systems and
         the Sun XI, ed. R. J. Garc\'{\i}a L\'opez, R. Rebolo, \&
         M. R. Zapatero Osorio (San Francisco: ASP), vol. 223, 1585 (B01)

\bibitem[Charbonnel et al.(1998)]{cbw98}
Charbonnel, C., Brown, J. A., \& Wallerstein, G. 1998, \aap, 332, 204

\bibitem[Che et al.(1983)]{chr83}
Che, A., Hempe, K., \& Reimers, D. 1983, \aap, 126, 225 (CHR83)

\bibitem[Cuntz (1990)]{c90}
Cuntz, M. 1990, \apj, 349, 141

\bibitem[Decin et al.(2003)]{detal03}
Decin, L., Vandenbussche, B., Waelkens, C., Decin, G., Eriksson, K.,
Gustafsson, B., Plez, B., \& Sauval, A. J. 2003, \aap, 400, 709 (DVW03)

\bibitem[de~Jager et al.(1988)]{jnh88}
de Jager, C., Nieuwenhuijzen, H., \& van der Hucht, K. A. 1988, 
\aaps, 72, 259 (dJNH88)

\bibitem[Drake (1985)]{d85}
Drake, S. A. 1985, in Progress in Stellar Spectral Line Formation Theory,
ed. J. E. Beckman \& L. Crivellari (Dordrecht: Reidel), 351

\bibitem[Drake \& Linsky (1983)]{dl83}
Drake, S. A., \& Linsky, J. L. 1983, \apj, 274, L77

\bibitem[Drake \& Linsky (1986)]{dl86}
Drake, S. A., \& Linsky, J. L. 1986, AJ, 91, 602 

\bibitem[Eggleton (1971)]{e71}
Eggleton, P. P. 1971, \mnras, 151, 351

\bibitem[Erhorn (1990)]{e90}
Erhorn, G. 1990, Ph.D. thesis, University of Hamburg

\bibitem[Griffin \& Lynas-Gray (1999)]{gl99}
Griffin, R. E. M., \& Lynas-Gray, A. E. 1999, AJ, 117, 2998 (GLG99)

\bibitem[Judge \& Stencel (1991)]{js91}
Judge, P. G., \& Stencel, R. E. 1991, \apj, 371, 357 (JS91)

\bibitem[Hagen et al.(1987)]{hhr87}
Hagen, H.-J., Hempe, K., \& Reimers, D. 1987, \aap, 184, 256 (HHR87)

\bibitem[Harper et al.(2001)]{hbl01}
Harper, G. M., Brown, A., \& Lim, J. 2001, \apj, 551, 1073 (HBL01)

\bibitem[Hartmann \& Avrett(1984)]{ha84}
Hartmann, L., \& Avrett, E. H. 1984, \apj, 284, 238

\bibitem[Hjellming \& Newell (1983)]{hn83}
Hjellming, R. M., \& Newell, R. T. 1983, \apj, 275, 704 (HN83)

\bibitem[Kudritzki \& Reimers (1978)]{kr78}
Kudritzki, R. P., \& Reimers, D. 1978, \aap, 70, 227 (KR78)

\bibitem[Lamers (1981)]{l81}
Lamers, H. J. G. L. M. 1981, \apj, 245, 593 (L81)

\bibitem[Levesque et al.(2005)]{letal05}
Levesque, E. M., Massey, P., Olsen, K. A. G., Plez, B., Josselin, E., 
          Maeder, A., \& Meynet, G. 2005, \apj, 628, 973 (LMO05)

\bibitem[Malmquist (1936)]{m36}
Malmquist, K. G. 1936, Stockholm Obs. Medd. No. 26 

\bibitem[Mauron \& Caux (1992)]{mc92}
Mauron, N., \& Caux, E. 1992, \aap, 265, 711

\bibitem[Musielak(2004)]{m04}
Musielak, Z. E. 2004, in Stars as Suns: Activity, Evolution and Planets,
IAU Symp. 219, ed. A. K. Dupree \& A. O. Benz (San Francisco: ASP), 437

\bibitem[Nieuwenhuijzen \& de~Jager (1990)]{nj90}
Nieuwenhuijzen, H., \& de Jager, C. 1990, \aap, 231, 134 (NdJ90)

\bibitem[Pols et al.(1997)]{petal97}
Pols, O. R., Tout, C. A., Schr\"oder, K.-P., Eggleton, P. P., \&
Manners, J. 1997, \mnras, 289, 869

\bibitem[Press et al.(1986)]{petal86}
Press, W. H., Flannery, B. P., Teukolsky, S. A., \& Vetterling, W. T.
1986, Numerical Recipies (Cambridge: Cambridge Univ. Press)

\bibitem[Reimers (1975)]{r75}
Reimers, D. 1975, Mem. Roy. Soc. Li\`ege 6. Ser. 8, 369 (R75)

\bibitem[Reimers (1977)]{r77}
Reimers, D. 1977, \aap, 61, 217 [Erratum: 67, 161] (R77)

\bibitem[Reimers \& Schr\"oder (1983)]{rs83}
Reimers, D., \& Schr\"oder, K.-P. 1983, \aap, 124, 241 (RS83)

\bibitem[Richichi \& Percheron (2002)]{rp02}
Richichi, A., \& Percheron, I. 2002, \aap, 386, 492

\bibitem[Schr\"oder (1983)]{s83}
Schr\"oder, K.-P. 1983, \aap, 124, L16

\bibitem[Schr\"oder et al.(1997)]{spe97}
Schr\"oder, K.-P., Pols, O. R., \& Eggleton, P. P. 1997, \mnras, 285,
696 (SPE97)

\bibitem[Schr\"oder et al.(1999)]{sws99}
Schr\"oder, K.-P., Winters, J. M., \& Sedlmayr, E. 1999, \aap, 349, 898

\bibitem[Schr\"oder \& Sedlmayr (2001)]{ss01}
Schr\"oder, K.-P., \& Sedlmayr, E. 2001, \aap, 366, 913

\bibitem[Schr\"oder et al.(2003)]{sww03}
Schr\"oder, K.-P., Wachter, A., \& Winters, J. M. 2003, \aap, 398, 229

\bibitem[Schr\"oder \& Pagel (2003)]{sp03}
Schr\"oder, K.-P., \& Pagel, B. E. J. 2003, \mnras, 343, 1231

\bibitem[Schr\"oder et al.(2004)]{spn04}
Schr\"oder, K.-P., Pauli, E.-M., \& Napiwotzki, R. 2004, \mnras, 354, 727

\bibitem[Schr\"oder \& Cuntz (2005)]{sc05}
Schr\"oder, K.-P., \& Cuntz, M. 2005, \apjl, 630, L73 (SC05) (Paper~I)

\bibitem[Taylor (1999)]{t99}
Taylor, B. J. 1999, \aaps, 134, 523

\bibitem[Thiering \& Reimers (1993)]{tr93}
Thiering, I., \& Reimers, D. 1993, \aap, 274, 838 (TR93)

\bibitem[van der Hucht et al.(1980)]{hbk80}
van der Hucht, K. A., Bernat A. P., \& Kondo Y. 1980, \aap, 82, 14

\bibitem[Wright (1970)]{w70}
Wright, K. O. 1970, Vistas in Astronomy, 12, 147

\end{thebibliography}
\end{document}